\documentclass{article}
\usepackage{smac}
\usepackage{times}
\usepackage{ifpdf}
\usepackage[english]{babel}
\usepackage{cite}
\usepackage{microtype}

\def\papertitle{musif: a Python package for symbolic music feature extraction}
\def\firstauthor{Ana Llorens}
\def\secondauthor{Second author}
\def\thirdauthor{Third author}

\newif\ifpdf
\ifx\pdfoutput\relax
\else
   \ifcase\pdfoutput
      \pdffalse
   \else
      \pdftrue
\fi

\ifpdf 
  \usepackage[pdftex,
    pdftitle={\papertitle},
    pdfauthor={\firstauthor, \secondauthor, \thirdauthor},
    bookmarksnumbered, 
    pdfstartview=XYZ 
   ]{hyperref}

  \usepackage[pdftex]{graphicx}
  \graphicspath{{./figures/}}
  \DeclareGraphicsExtensions{.pdf,.jpeg,.png}

  \usepackage[figure,table]{hypcap}

\else 
  \usepackage[dvips,
    bookmarksnumbered, 
    pdfstartview=XYZ 
  ]{hyperref}  

  \usepackage[dvips]{epsfig,graphicx}
  \graphicspath{{./figures/}}
  \DeclareGraphicsExtensions{.eps}

  \usepackage[figure,table]{hypcap}
\fi

\hypersetup{
    colorlinks,%
    citecolor=black,%
    filecolor=black,%
    linkcolor=black,%
    urlcolor=black
}

\title{\papertitle}

\oneauthor
  {
    Ana Llorens,\textsuperscript{1}
    Federico Simonetta,\textsuperscript{2}
    Mart\'in Serrano,\textsuperscript{2}
    \'Alvaro Torrente\textsuperscript{1, 2}
  } {
    \textbf{1} Department of Musicology, Universidad Complutense de Madrid, Madrid, Spain  \\
    \textbf{2} Instituto Complutense de Ciencias Musicales, Madrid, Spain \\ 
    {\tt {\{first letter of the first name\}\{surname\}@iccmu.es}
    }
}

\usepackage[frozencache,cachedir=.]{minted}
\usepackage{stfloats}
\usepackage{paralist}

\begin{document}
\capstartfalse
\maketitle
\capstarttrue

\begin{abstract}
In this work, we introduce \texttt{musif}, a Python package that facilitates the automatic extraction of features from symbolic music scores. The package includes the implementation of a large number of features, which have been developed by a team of experts in musicology, music theory, statistics, and computer science. Additionally, the package allows for the easy creation of custom features using commonly available Python libraries. \texttt{musif} is primarily geared towards processing high-quality musicological data encoded in MusicXML format, but also supports other formats commonly used in music information retrieval tasks, including MIDI, MEI, Kern, and others. We provide comprehensive documentation and tutorials to aid in the extension of the framework and to facilitate the introduction of new and inexperienced users to its usage.
\end{abstract}

\section{Introduction}\label{sec:introduction}

The abstraction represented in music scores, which are symbolic representations of music, has been shown to be highly relevant for both cognitive and musicological studies. In cognitive studies, the abstraction process used by human music cognition to categorize sound is important to understand how we identify and perceive different musical aspects, such as timbres, pitches, durations, and rhythms~\cite{deutsch2013PsychologyMusic}. In musicological studies, the abstraction represented in music scores is important as it provides a direct source of information to understand how the music was constructed. Throughout history, these aspects have been encoded in different forms, with common Western music notation being the most widely used in the Western world for centuries. Therefore, music notation is considered of paramount importance in the field of musicology.

In the field of sound and music computing, however, research has primarily focused on analyzing music in the audio domain, while other modalities such as images and scores have received less attention~\cite{simonetta2019MultimodalMusicInformation}. Researchers interested in applying machine learning methods to the analysis of music scores will likely seek methods for representing them in a suitable way. In the context of modern deep learning and machine learning, two main approaches have emerged: feature learning~\cite{bengio2013RepresentationLearningReview} and feature extraction~\cite{cuthbert2011feature,mckay2018jsymbolic}. Feature learning -- or representation learning -- involves using algorithms to learn the features from the data in a way that is optimal to the specific statistical inference problem and is mainly applied with Neural Networks~\cite{simonetta2019convolutional,prang2020SignaldomainRepresentationSymbolic,lisena2022MIDI2vecLearningMIDI}; feature extraction,  instead, involves the computation of generic and hand-crafted features, needing further successive steps such as feature selection and dimensionality reduction. Both approaches have their own set of advantages and disadvantages and the choice of which approach to use will depend on the specific task and the available data. Here we focus on the latter exclusively.

Feature extraction has widely been used in various machine learning tasks and has been partially successful in music computing~\cite{bigo2017SketchingSonataForm,kempfert2020where,simonetta2022PerceptualMeasureEvaluating}. However, a major drawback is the effort and time required to craft useful features for a specific task. To address this issue, researchers have previously proposed software tools that assist in extracting features from music, such as audio files and scores. Additionally, with the advancement of modern computer languages such as Python and JavaScript, the implementation of new features has become easier and more accessible.

Musicologists may also resort to feature extraction, especially in the context of the so-called corpus studies. In fact, existing software for symbolic music feature extraction -- e.g.\  jSymbolic~\cite{mckay2018jsymbolic} -- was partly designed to help musicologists obtain the data they required in a fast and accurate way. This is especially important because the computation of the features could hardly be achieved by the manual work of musicologists, who, as of today, devote time to manual annotations such as harmony~\cite{neuwirth2018AnnotatedBeethovenCorpus} and cadence~\cite{hentschel2021AnnotatedMozartSonatas,romcadence}. Examples of such feature-driven, computational musicology can be found in studies of musical form~\cite{choro}, harmony~\cite{hentschelchords}, and compositional styles~\cite{anuario,cuenca,Carestini,Berlin,Rordriguez}, among others.

In this work, we introduce a software tool named \texttt{musif}, which offers a comprehensive collection of features that are extracted from various file formats. The tool is designed to be easily extensible using the Python programming language and is specifically tailored for 18th-century opera arias, although it has been tested on a variety of other repertoires, including Renaissance and Pop music. Furthermore, in contrast to previous software~\cite{mckay2018jsymbolic,cuthbert2011feature}, \texttt{musif} is developed with a focus on musicological studies and is thus geared towards high-quality music datasets, addressing the issue of limited data availability that is commonly encountered in feature learning methods.

To aid in its usage, \texttt{musif} is accompanied by detailed software documentation \footnote{\url{https://musif.didone.eu}}. This documentation provides adequate information for both novice and advanced users, enabling them to take full advantage of the tool and add new features and file formats as needed. The project is developed using open source methods and adopts GitHub to manage issues and pull requests, as well as to distribute the source code\footnote{\url{https://github.com/DIDONEproject/musif}}. 

\begin{figure*}
    \centering
    \includegraphics[width=\textwidth]{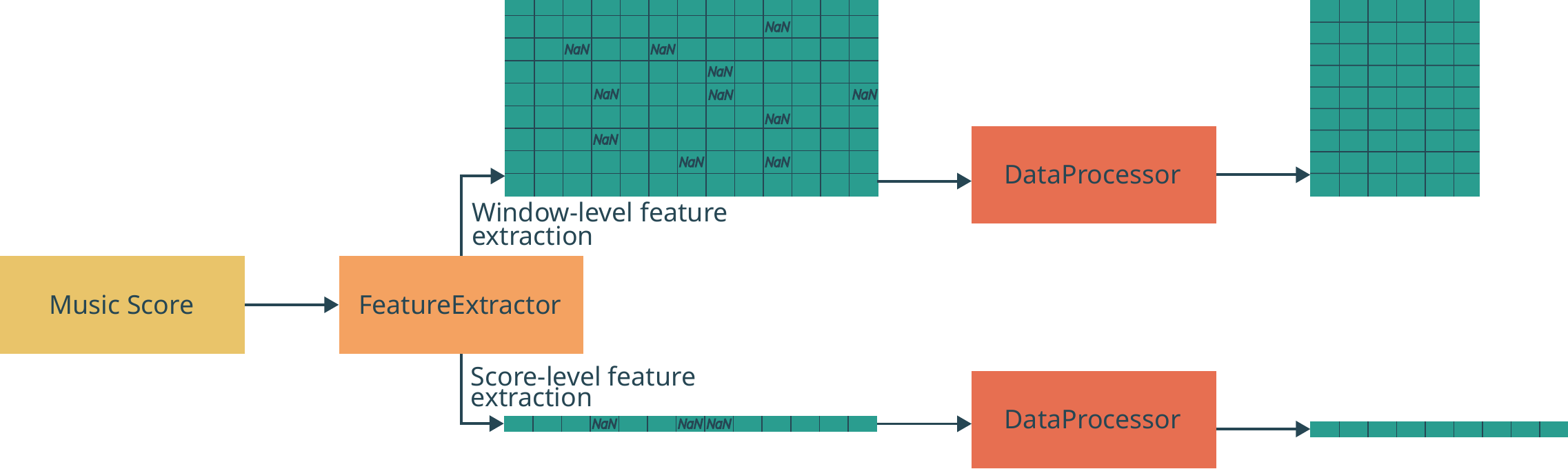}
    \caption{Flow chart of the general pipeline for feature extraction with \texttt{musif} from a single score. First, features are extracted from a music score. If window-level features are used, then a row for each window of measures is generated, otherwise a single row for the whole score. Then, the \texttt{DataProcessor} cleans the table by removing NaN, replacing them with 0, and merging or removing the undesired columns.}
    \label{fig:general_pipeline}
    \hfill
\end{figure*}

\hfill
\section{Design Principles}
\label{sec:principles}
\hfill

The development of the \texttt{musif} was guided by four key design principles. 

The foremost principle was the ability to customize and extend the framework to meet the user's specific requirements. This includes the capability to alter the feature extraction process by introducing new features coded by the user and by modifying the existing pipeline.

The second principle that guided the development was to ensure the usability of the software by individuals with minimal technical expertise, with musicologists as the primary target audience. This principle mainly entailed providing a user-friendly interface for the entire feature extraction process, with default settings that are deemed optimal. Additionally, comprehensive documentation was produced to aid novice users in understanding the feature extraction process of symbolic music.

As musicologists were identified as the primary target audience, special attention was paid to the file types supported by the system. Specifically, an effort was made to find a combination of file formats that were both easy to create and able to represent musicological annotations, which could be used as sources for feature extraction.

The final principle that underpins the entire structure of \texttt{musif} is its suitability for big data analysis. Specifically, measures were taken to ensure that the framework was computationally efficient on commercially available computers.

\newpage
\section{Implementation}
\label{sec:implementation}

\begin{listing*}
    \begin{minted}[frame=single,framesep=10pt,breaklines,breakafter=d,fontsize=\small]{python}
from musif.extract.extract import FeaturesExtractor
from musif.process.processor import DataProcessor

features = FeaturesExtractor(
    # here we use `None`, but it could be the path to a YAML file containing 
    # specifications
    None,
    # the options below override the YAML file if it is provided
    xml_dir="data_notation",
    musescore_dir="data_harmony",
    basic_modules=["scoring"],
    features=["core", "ambitus", "interval", "tempo", 
              "density", "texture", "lyrics", "scale", 
              "key", "dynamics", "rhythm"]
).extract()

# For the DataProcessor, the arguments are the extracted table and the path to a YAML 
# file
# As before, the YAML file can be overridden by variadic arguments
processed_features = DataProcessor(features, None).process().data

# the output is a pandas DataFrame!
    \end{minted}
    \caption{Example of usage feature extraction with default options and stock features}
    \label{code:1}
\end{listing*}

\subsection{General pipeline}

The implementation of \texttt{musif} is mainly based on music21~\cite{cuthbert2011feature} and methodically divided into two primary stages, both of which are highly configurable. Fig.~\ref{fig:general_pipeline} shows a flowchart of the general pipeline. 

The initial stage pertains to the actual extraction of features, during which a substantial number of features are derived from the data. Among these features, some are solely designed for the calculation of ``second-order'' features, which are derived from the primary ones. For instance, the number of notes on a score may not hold inherent significance, but it acquires meaning when considered in relation to the total length of the score. Therefore, an additional operation is required to compute the ratio between the number of notes and features that denote the total duration of the score, such as the total number of beats. As a result of this, certain "first-order" features may not be relevant for the specific task at hand.

To address this issue, we have implemented an additional step that we refer to as ``post-processing''. In this stage, certain ``first-order'' features are eliminated, while others are aggregated according to the user specifications. For example, to lower the overall number of features and attain a more succinct representation, the user may choose to aggregate features that originate from similar instruments, such as strings, by utilizing statistical measures such as the mean, the variance, and other statistical moments. Another crucial task accomplished during post-processing is the standardization of representation for missing data, such as NaN values or empty strings.

The aforementioned two steps correspond to two Python objects, namely, the \texttt{FeaturesExtractor} and the\\\texttt{DataProcessor}. Both of these objects take as input an extensible configuration, which can be expressed in various ways, namely variadic python arguments in the class constructor and/or a YAML file. The configuration of the \texttt{FeaturesExtractor} object includes the path to the data, the features that should be extracted, the paths or objects containing custom features, and other similar requirements. For its part, the configuration of the\\\texttt{DataProcessor} object offers the flexibility to specify the columns that should be aggregated or removed, as well as the columns in which NaN values should be replaced with a default value, such as zero.

\begin{figure*}
    \centering
    \includegraphics[width=\textwidth]{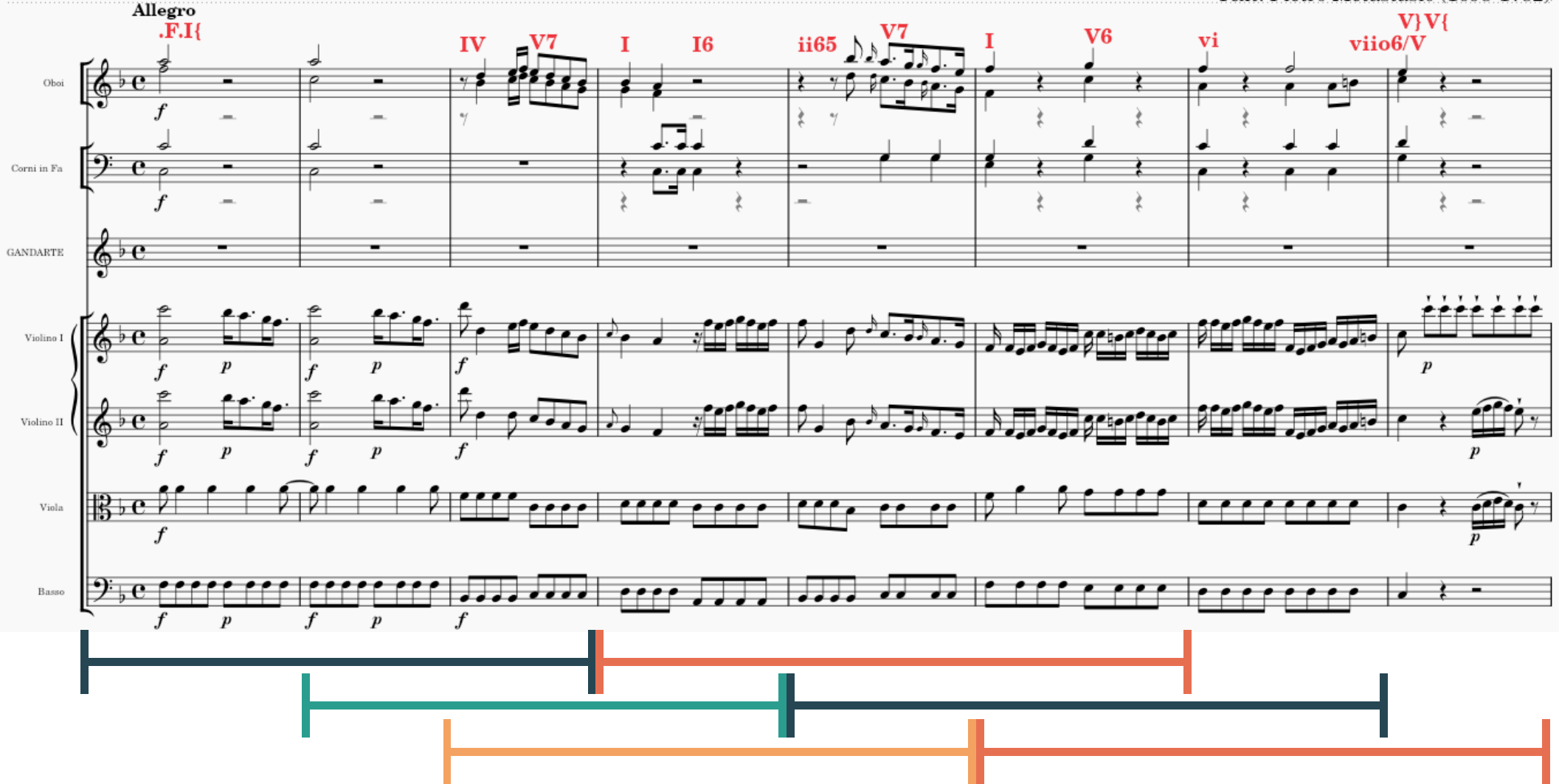}
    \caption{Example of windowing on a music score. The unit of measure for the window length and overlap are measures. In this case, windows have a length of 3 and an overlap of 2. At the top of the score, in red, an example of harmonic analysis is shown.}
    \label{fig:windows}
\end{figure*}

The outcome of the entire process is a tabular representation, with one column per feature and one row per musical score. Optionally, scores can be analyzed using moving windows, in which case the output table will have one row for each window. When using windows, the window size and overlap can be specified as the number of measures, as shown in Fig.~\ref{fig:windows}. A sample code that demonstrates the usage of the tool is provided in Listing~\ref{code:1}.

\subsection{File formats}

Given that our primary objective was to develop a software tool for musicological applications, it was imperative to support file formats that are easily usable in musicological analysis. As such, we carefully considered file formats such as MusicXML, MEI, and IEEE 1599. These file formats can represent common Western music notation with a high degree of detail and have been utilized for both musicological and MIR tasks. However, it was determined that only MusicXML is fully supported by user-end graphical interfaces. The requirement for users to possess both musicological training and the ability to effectively utilize advanced software for editing large XML files is a rare combination, and, as such, it was not deemed a viable inclusion in the design of the system. Moreover, certain features implemented by \texttt{musif} are derived from functional, Roman-numeral harmonic analysis, which cannot be represented in the standard format of MusicXML. To solve this issue, we have adopted the MuseScore file format, in line with previous works in this field~\cite{neuwirth2018AnnotatedBeethovenCorpus,hentschel2021AnnotatedMozartSonatas}.

Overall, the recommended file formats for the \texttt{musif} system are MusicXML for notation parsing and MuseScore for harmonic annotations. However, if only MuseScore files are available, the \texttt{MuseScore} software can be utilized to generate the necessary MusicXML files. Additionally, alternative file formats may be employed in place of MusicXML by leveraging the \texttt{music21} library for parsing notation files. This approach supports a comprehensive array of file formats. Furthermore, any file format supported by MuseScore can be utilized through automatic conversion to MusicXML. This pipeline is particularly recommended for extracting features from MIDI files.

However, the parsing approach adopted in this system may be relatively slow when working with a large number of files. To mitigate this issue, a caching system has been implemented in order to save to disk any property, function, or method result that originates from \texttt{music21} objects. This approach has been tested and has demonstrated a significant improvement in processing speed, with a 2 to 3 times increase in speed observed when cached files are used. This caching system is particularly useful when designing or debugging feature extractions on a large number of files, as it allows for more efficient and expedient processing.

\subsection{Customization}

To facilitate customization of the feature extraction process, three main tools are available. These tools allow for more flexibility and precision in the feature extraction process, enabling users to tailor the process to their specific needs and requirements. These tools are further described in the subsequent list.

\begin{enumerate}
    \item \textit{Custom features}: The user can add custom features by developing two simple functions: one to extract features from each individual part in the score, and another to extract features from the entire score. This second function can optionally utilize the features extracted from the individual parts. Additionally, the user can specify the extraction order and feature dependencies, allowing for the use of previously extracted features in the computation of newer features. The implementation of these custom features can be easily accomplished using the \texttt{music21} Python library.
    
    \item \textit{Hooks}: Hooks are user-provided functions that are called at specific stages of the extraction process. In the current version of \texttt{musif}, only one type of hook is possible, namely just after the parsing of the input files is completed and just before the caching mechanism is initialized. The user can provide a list of functions that accept the parsed score as input and that are run before the caching mechanism is initialized. When using the cached files, these hooks will no longer be run. This hook is particularly useful for modifying the input scores before caching, such as deleting or modifying unsupported notation elements from \texttt{music21} objects, thus mitigating the constraints of the caching mechanism, which only allows read-only operations on the scores.

    \item \textit{Python mechanisms}: The Python programming language offers a range of advanced methods for modifying and extending existing software. As \texttt{musif} is fully implemented in pure Python, these methods are fully applicable. They include, but are not limited to, class inheritance, method and property overriding, and type casting.
    
\end{enumerate}

\section{Stock Features}
\label{sec:features}

\texttt{musif} is distributed with a wide variety of features already implemented. These sets of features can be selected for extraction using the \texttt{FeaturesExtractor}'s constructor arguments -- see Listing~\ref{code:1} --, while the \texttt{DataProcessor} can be utilized for further refining the desired features. Each set corresponds to a specific Python sub-package. The total number of features varies based on the instrumentation used in the score and is usually between 500 feature values for simple monophonic scores and more than 10,000 feature values for orchestral scores. In this presentation, we will provide a brief summary of each of these modules. For those who wish to carefully select features, more detailed information can be found in the online documentation, including pre-made Python regular expressions that can be used to easily select the desired features.

In general, all the features were designed to be meaningful for musicologists and music theorists, giving value to studies attempting to explain statistical results on the basis of the features. The modular structure of the features also allows researchers to conveniently focus their analysis on only certain aspects of the music.

Here, we will use the word \textit{sound} to refer to a specific timbre -- e.g.\ violin -- which can be repeated multiple times in the score -- e.g.\ violin I and violin II. Moreover, we will use \textit{family} to refer to a family of instruments -- e.g.\ strings, voices, brasses, and so on. The stock feature modules available in \texttt{musif} are as follows:

\begin{itemize}

    \item \textit{Core}: These features are essential for the identification of music scores and for subsequent elaboration. They are always required and include the total number of measures and notes, as well as the number of measures containing notes and their averages for each sound or part and for each family and/or score. Other examples of such features include the filename of the score, the time signature, and the key signature.
    
    \item \textit{Scoring}: This module computes features that are related to the instrumentation and voices used in the score. Examples of features in this module include the instruments, families, and parts present in the score, as well as the number of parts for each instrument and family in the score. This module can be used to get a better understanding of the orchestration used in the composition.

    \item \textit{Key}: This module computes features that are related to the key signature and tonality, i.e., the key, of the piece. Examples of features in this module include the Krumhansl-Schmuckler tonality estimation~\cite{krumhansl1990CognitiveFoundationsMusical}, the key signature, and the mode (major or minor). This module allows for analyzing the underlying tonal system used in the composition.

    \item \textit{Tempo}: This module computes features that are related to the tempo marking on the score. It should be noted that since some features depend on the terminology used by the composer for the tempo indication, some of these features may not be reliable for all repertoires. In fact, as the composers' marking need not be expressed quantitatively -- it is actually more typical in some repertoires to have just a verbal indication -- the numerical values extracted by \texttt{musif} ultimately depend on the BPM value given during the engraving process, if available.

    \item \textit{Density}: These features relate the number of notes with respect to the total number of measures, as well as with respect to the total number of measures that contain sound, for a single part, sound, or family. This module provides insights into the density of the sound in the composition and allows comparing the activity level of different parts or families in the score.

    \item \textit{Harmony}: This is one of the largest feature modules; it computes features based on the harmonic annotations provided in the MuseScore files according to a previous standard~\cite{neuwirth2018AnnotatedBeethovenCorpus,hentschel2021AnnotatedMozartSonatas}. Examples of these features include the number of harmonic annotations, the number of chords performing the tonic, dominant, and sub-dominant functions, the harmonic rhythm -- i.e. the rate of harmonic changes in relation to the number of beats or measures --, as well as features related to modulations annotated in the MuseScore files. This module can be used to get a better understanding of the harmonic structure of the composition and to analyze the harmonic progressions used in the composition.

    \item \textit{Rhythm}: This module computes features related to the note durations and to particular rhythmic figures, such as dotted and double-dotted rhythms. Examples of features in this module include the average note duration and the frequency of particular rhythmic figures. This module analyzes the rhythmic structure of the composition and the rhythmic patterns used in it.

    \item \textit{Scale}: This module computes features related to specific melodic degrees with respect to the main key of the score, as computed in the \textit{key} module, and to the local key, as provided in the MuseScore harmonic annotations. Examples of features in this module include the frequency of specific scale degrees in a given part.

    \item \textit{Dynamics}: This module computes features related to the distribution of dynamic markings across the score, by assigning numerical values to each dynamic marking according to their corresponding intensity.  As is the case with tempo, the specific numerical value of a given dynamic marking is assigned during the engraving process, with some software assigning default values that the engraver may need to modify depending on the notation conventions. Similarly to other features, this module may not be completely generalizable to some repertoires, as the interpretation of dynamic markings can vary across different compositions and styles, or even be completely absent. Examples of features in this module include the frequency of specific dynamic markings, the average dynamic level, and the distribution of dynamic markings across the score. This module extracts information about the expressivity of the composition and analyzes the use of dynamic contrasts in it.

    \item \textit{Ambitus}: This module computes the ambitus, or melodic range, of the piece in semitones, for the whole piece as well as for each individual part, sound, or family. It also computes the lowest and highest pitches and the note names thereof. 

    \item \textit{Melody}: This module computes an extensive number of features related to the distribution and types of melodic intervals for each part, voice, sound, and family. This is the largest set of features within \texttt{musif}. Examples of features in this module include the frequency of specific interval types, the distribution of interval sizes, and the proportion of ascending and descending intervals. This module provides insights into the melodic structure of the composition by analyzing the use of specific intervals in it.

    \item \textit{Lyrics}: This module considers the alignment between lyrics, if available, and the notes and computes features related to their distribution. Examples of features in this module include the total number of syllables in each vocal part, the average number of notes per syllable, and the proportion of measures that contain notes for each vocal part in the score. This module can facilitate a more profound comprehension of the relationships between lyrics and music in the composition.

   \item \textit{Texture}: This module computes the ratio of the number of notes between two parts, considering all the possible pairs of parts. This feature can provide insight into the relative density and activity level of different parts in the score and can be used to analyze the texture of the composition.
    
\end{itemize}

\section{Discussion and Future Works}
\label{sec:discussion}

This work presents the \texttt{musif} module to the scientific community as a tool for the extraction of features from symbolic music scores. It is designed with a focus on extensibility and customization, while also providing good defaults for the novice user and supporting musicologically-curated datasets. The module is implemented in Python, and it provides a comprehensive set of features covering various aspects of music scores, including harmony, rhythm, melody, and many more. The modular structure of the \texttt{musif} makes it easy to use and customize according to the user's needs. 

In comparison to existing software such as \texttt{jsymbolic}~\cite{mckay2018jsymbolic} and \texttt{music21}~\cite{cuthbert2011feature}, \texttt{musif} offers a significantly larger number of features, approximately 2 times larger. Additionally, \texttt{jsymbolic} computes features based on pure MIDI encoding, with only 2 features based on the MEI format. This is an essential aspect for musicological studies as MIDI, although commonly used in the MIP field, is not capable of representing various characteristics of music notation, such as alterations, key signatures, rhythmic and dynamic annotations, chords, and lyrics.

\texttt{music21} already implements several features based on its powerful parsing engine, which allows it to take full advantage of MusicXML, MEI, and Kern features. However, \texttt{musif} expands upon this set of computable features while remaining completely based on \texttt{music21} and allowing the automatic extraction of features at the window level. Furthermore, it includes a caching system that allows for improved performance during the feature extraction process. This caching system saves the results of computations to disk, reducing the need to perform the same calculations multiple times, thus making the extraction process more efficient. Thus, \texttt{musif} provides a more extensive set of features while being highly performant in its extraction process, making it a valuable tool for researchers in the field of music information retrieval and musicology.

While this paper describes the release of \texttt{musif} 1.0, we are aware that there is wide room to improve \texttt{musif} further, making it faster, more general, usable, and accurate. Specifically, we want to improve three aspects of the software:

\begin{itemize}

    \item \textit{Data visualization}: we want to provide the user with tools that help the visualization of the data that \texttt{musif} extracts; this aspect would particularly be useful for preliminary analysis.

    \item \textit{Repertoire}: As of now, \texttt{musif} has been tested on other types of corpora for different music styles, including EWLD\~cite{simonetta2018symbolic}, Humdrum database~\cite{sapp2005online}, piano scores and performances~\cite{foscarin2020asap}, and masses from the Renaissance~\cite{cumming2018methodologies}. It has additionally been utilized on an in-house corpus of more than 1600 opera arias. For this reason, most of the design choices and of the implemented features target this repertoire. We want to make it more powerful and efficient for other repertoires too.

    \item \textit{More numerical features}: Although \texttt{musif} already provides a wide set of musical features, we are sure that many other features could be defined and included in \texttt{musif}, empowering both musicological analysis and data science studies.

\end{itemize}

We also plan to study in more depth the comparison between the existing tools for music feature extraction, including benchmarks and test performances.

While we continue working on these paths, we hope that \texttt{musif} can be a valuable tool for the Sound and Music Computing community and welcome any suggestions or contributions to the software. We encourage the community to use and test \texttt{musif} and provide feedback so that we can continue to improve and develop it further. It is our goal to make \texttt{musif} a widely used and reliable tool for MIP and musicology research.

\begin{acknowledgments}
This work is a result of the Didone Project\cite{didone}, which has received funding from the European Research Council (ERC) under the European Union’s Horizon 2020 research and innovation program, Grant agreement No. 788986. It has also been conducted with funding from Spain’s Ministry of Science and Innovation (\href{https://doi.org/10.13039/501100011033}{IJC2020-\-043969-I/\-AEI/\-10.13039/\-501100011033}).
\end{acknowledgments} 

\bibliography{musif,bibliography}

\begin{thebibliography}{10}
\providecommand{\url}[1]{#1}
\csname url@samestyle\endcsname
\providecommand{\newblock}{\relax}
\providecommand{\bibinfo}[2]{#2}
\providecommand{\BIBentrySTDinterwordspacing}{\spaceskip=0pt\relax}
\providecommand{\BIBentryALTinterwordstretchfactor}{4}
\providecommand{\BIBentryALTinterwordspacing}{\spaceskip=\fontdimen2\font plus
\BIBentryALTinterwordstretchfactor\fontdimen3\font minus
  \fontdimen4\font\relax}
\providecommand{\BIBforeignlanguage}[2]{{%
\expandafter\ifx\csname l@#1\endcsname\relax
\typeout{** WARNING: IEEEtran.bst: No hyphenation pattern has been}%
\typeout{** loaded for the language `#1'. Using the pattern for}%
\typeout{** the default language instead.}%
\else
\language=\csname l@#1\endcsname
\fi
#2}}
\providecommand{\BIBdecl}{\relax}
\BIBdecl

\bibitem{deutsch2013PsychologyMusic}
D.~Deutsch, \emph{The {{Psychology}} of {{Music}}}, 3rd~ed.\hskip 1em plus
  0.5em minus 0.4em\relax {Academic Press}, 2013.

\bibitem{simonetta2019MultimodalMusicInformation}
F.~Simonetta, S.~Ntalampiras, and F.~Avanzini, ``Multimodal {{Music Information
  Processing}} and {{Retrieval}}: {{Survey}} and {{Future Challenges}},'' in
  \emph{Proceedings of 2019 {{International Workshop}} on {{Multilayer Music
  Representation}} and {{Processing}}}.\hskip 1em plus 0.5em minus 0.4em\relax
  {Milan, Italy}: {IEEE Conference Publishing Services}, 2019, pp. 10--18. doi:
  \href{https://doi.org/10.1109/mmrp.2019.00012}{10.1109/mmrp.2019.00012}

\bibitem{bengio2013RepresentationLearningReview}
Y.~Bengio, A.~Courville, and P.~Vincent, ``Representation {{Learning}}: {{A
  Review}} and {{New Perspectives}},'' \emph{IEEE Transactions on Pattern
  Analysis and Machine Intelligence}, vol.~35, no.~8, pp. 1798--1828, Aug.
  2013. doi:
  \href{https://doi.org/10.1109/TPAMI.2013.50}{10.1109/TPAMI.2013.50}

\bibitem{cuthbert2011feature}
M.~S. Cuthbert, C.~Ariza, and L.~Friedland, ``Feature extraction and machine
  learning on symbolic music using the music21 toolkit,'' in \emph{Proceedings
  of the 12th International Society for Music Information Retrieval Conference,
  {{ISMIR}} 2011, Miami, Florida, {{USA}}, October 24-28, 2011}.\hskip 1em plus
  0.5em minus 0.4em\relax {University of Miami}, 2011, pp. 387--392. doi:
  \href{https://doi.org/10.5281/zenodo.1416288}{10.5281/zenodo.1416288}

\bibitem{mckay2018jsymbolic}
C.~McKay, J.~Cumming, and I.~Fujinaga, ``{{jSymbolic}} 2.2: {{Extracting}}
  features from symbolic music for use in musicological and {{MIR}} research,''
  in \emph{Proceedings of the 19th {{International Society}} for {{Music
  Information Retrieval Conference}}}, 2018. ISBN 978-2-9540351-2-3 pp.
  348--354. doi:
  \href{https://doi.org/10.5281/zenodo.1492421}{10.5281/zenodo.1492421}

\bibitem{simonetta2019convolutional}
F.~Simonetta, C.~E. {Cancino-Chac{\'o}n}, S.~Ntalampiras, and G.~Widmer, ``A
  convolutional approach to melody line identification in symbolic scores,'' in
  \emph{Proceedings of the 20th International Society for Music Information
  Retrieval Conference}.\hskip 1em plus 0.5em minus 0.4em\relax {Delft, The
  Netherlands}: {ISMIR}, Nov. 2019, pp. 924--931. doi:
  \href{https://doi.org/10.5281/zenodo.3527966}{10.5281/zenodo.3527966}

\bibitem{prang2020SignaldomainRepresentationSymbolic}
M.~Prang and P.~Esling, ``Signal-domain representation of symbolic music for
  learning embedding spaces,'' {Stockholm, Sweden}, p.~10, Oct. 2020.

\bibitem{lisena2022MIDI2vecLearningMIDI}
P.~Lisena, A.~{Mero{\~n}o-Pe{\~n}uela}, and R.~Troncy, ``{{MIDI2vec}}:
  {{Learning MIDI}} embeddings for reliable prediction of symbolic music
  metadata,'' \emph{Semantic Web}, vol.~13, no.~3, pp. 357--377, Jan. 2022.
  doi: \href{https://doi.org/10.3233/SW-210446}{10.3233/SW-210446}

\bibitem{bigo2017SketchingSonataForm}
L.~Bigo, M.~Giraud, R.~Groult, N.~{Guiomard-Kagan}, and F.~Lev{\'e},
  ``Sketching sonata form structure in selected classical string quartets.'' in
  \emph{Proceedings of the 18th International Society for Music Information
  Retrieval Conference}.\hskip 1em plus 0.5em minus 0.4em\relax {Suzhou,
  China}: {ISMIR}, Oct. 2017, pp. 752--759. doi:
  \href{https://doi.org/10.5281/zenodo.1415020}{10.5281/zenodo.1415020}

\bibitem{kempfert2020where}
K.~C. Kempfert and S.~W.~K. Wong, ``Where does {{Haydn}} end and {{Mozart}}
  begin? {{Composer}} classification of string quartets,'' \emph{Journal of New
  Music Research}, vol.~49, no.~5, pp. 457--476, Oct. 2020. doi:
  \href{https://doi.org/10.1080/09298215.2020.1814822}{10.1080/09298215.2020.1814822}

\bibitem{simonetta2022PerceptualMeasureEvaluating}
F.~Simonetta, F.~Avanzini, and S.~Ntalampiras, ``A {{Perceptual Measure}} for
  {{Evaluating}} the {{Resynthesis}} of {{Automatic Music Transcriptions}},''
  \emph{Multimedia Tools and Applications}, 2022. doi:
  \href{https://doi.org/10.1007/s11042-022-12476-0}{10.1007/s11042-022-12476-0}

\bibitem{neuwirth2018AnnotatedBeethovenCorpus}
M.~Neuwirth, D.~Harasim, F.~C. Moss, and M.~Rohrmeier, ``The {{Annotated
  Beethoven Corpus}} ({{ABC}}): {{A Dataset}} of {{Harmonic Analyses}} of {{All
  Beethoven String Quartets}},'' \emph{Frontiers in Digital Humanities},
  vol.~5, 2018. doi:
  \href{https://doi.org/10.3389/fdigh.2018.00016}{10.3389/fdigh.2018.00016}

\bibitem{hentschel2021AnnotatedMozartSonatas}
J.~Hentschel, M.~Neuwirth, and M.~Rohrmeier, ``The {{Annotated Mozart
  Sonatas}}: {{Score}}, {{Harmony}}, and {{Cadence}},'' \emph{Transactions of
  the International Society for Music Information Retrieval}, vol.~4, no.~1,
  pp. 67--80, May 2021. doi:
  \href{https://doi.org/10.5334/tismir.63}{10.5334/tismir.63}

\bibitem{romcadence}
O.~Raz, D.~Chawin, and U.~B. Rom, ``{The Mozart Expositional Punctuation
  Corpus: A Dataset of Interthematic Cadences in Mozart's Sonata-Allegro
  Expositions},'' \emph{Empirical Musicology Review}, vol.~16, pp. 134--144,
  2021. doi: \href{https://doi.org/10/grq2fp}{10/grq2fp}

\bibitem{choro}
F.~Moss, W.~Fernandes~de Souza, and M.~Rohrmeier, ``{Harmony and Form in
  Brazilian Choro: A Corpus-Driven Approach to Musical Style Analysis},''
  \emph{Journal of New Music Research}, vol.~49, pp. 416--437, 2020. doi:
  \href{https://doi.org/10/grq2fm}{10/grq2fm}

\bibitem{hentschelchords}
J.~Hentschel, F.~C. Moss, A.~McLeod, M.~Neuwirth, and M.~Rohrmeir, ``{Towards a
  Unified Model of Chords in Western Harmony},'' in \emph{{Music Encoding
  Conference 2021}}, 2022, pp. 143--149. doi:
  \href{https://doi.org/10/grq2fk}{10/grq2fk}

\bibitem{anuario}
A.~Llorens and A.~Torrente, ``{Constructing \textit{opera seria} in the Iberian
  Courts: Metastasian Repertoire for Spain and Portugal},'' \emph{Anuario
  Musical}, vol.~76, pp. 73--110, Jul. 2021. doi:
  \href{https://doi.org/10/grq2fn}{10/grq2fn}

\bibitem{cuenca}
M.~E. Cuenca and C.~McKay, ``{Exploring Musical Style in the Anonymous and
  Doubtfully Attributed Mass Movements of the Coimbra Manuscripts: A
  Statistical Approach},'' in \emph{{Medieval and Renaissance Music
  Conference}}, 2019.

\bibitem{Carestini}
V.~Anzani and A.~Llorens, ``{Shaping Eighteenth-Century Opera: The Singer's
  Impact},'' in \emph{{Tosc@ Junior Conference}}, 2021.

\bibitem{Berlin}
A.~Torrente, ``{\textit{Didone trasmutata}: Aria Settings and the Expression of
  Emotions in Metastasian Operas},'' in \emph{{Mapping Artistic Networks of
  Italian Theatre and Opera across Europe, 1600--1800}}, 2019.

\bibitem{Rordriguez}
E.~Rodriguez-Garcia and C.~McKay, ``{\textit{Ave festiva ferculis}: Exploring
  Attribution by Combining Manual and Computational Analysis.}'' in
  \emph{{Medieval and Renaissance Music Conference}}, 2021.

\bibitem{krumhansl1990CognitiveFoundationsMusical}
C.~L. Krumhansl, \emph{Cognitive {{Foundations}} of {{Musical Pitch}}}.\hskip
  1em plus 0.5em minus 0.4em\relax {Oxford University Press}, 1990. ISBN
  0-19-505475-X

\bibitem{sapp2005online}
C.~S. Sapp, ``Online {{Database}} of {{Scores}} in the {{Humdrum File
  Format}},'' in \emph{Proceedings of the 6th {{International Conference}} on
  {{Music Information Retrieval}}}, 2005, p.~2. doi:
  \href{https://doi.org/10.5281/zenodo.1417281}{10.5281/zenodo.1417281}

\bibitem{foscarin2020asap}
F.~Foscarin, A.~Mcleod, P.~Rigaux, F.~Jacquemard, and M.~Sakai, ``{{ASAP}}: A
  dataset of aligned scores and performances for piano transcription,'' in
  \emph{Proceedings of the 21st {{International Society}} for {{Music
  Information Retrieval}}}, 2020, Proceedings. doi:
  \href{https://doi.org/10.5281/zenodo.4245489}{10.5281/zenodo.4245489}

\bibitem{cumming2018methodologies}
J.~Cumming, C.~McKay, J.~Stuchbery, and I.~Fujinaga, ``Methodologies for
  {{Creating Symbolic Corpora}} of {{Western Music Before}} 1600,'' in
  \emph{Proceedings of the {{ISMIR}}}.\hskip 1em plus 0.5em minus 0.4em\relax
  {Paris, France}: {ISMIR}, Sep. 2018, pp. 491--498. doi:
  \href{https://doi.org/10.5281/zenodo.1492459}{10.5281/zenodo.1492459}

\bibitem{didone}
A.~Torrente and A.~Llorens, ``{The Musicology Lab: Teamwork and the
  Musicological Toolbox},'' in \emph{{Music Encoding Conference 2021}}, 2022,
  pp. 9--20. doi: \href{https://doi.org/10/grqp2b}{10/grqp2b}

\end{thebibliography}

\end{document}